# Beyond Topicality: A Conceptual Analysis of Societal Relevance and Its Application to Search Results and AI Responses

**Lewandowski, Dirk**

Hamburg University of Applied Sciences, Germany | dirk.lewandowski@haw-hamburg.de

## ABSTRACT

This paper examines "societal relevance," a concept introduced by Haider and Sundin to address the limitations of traditional relevance models in web search. While topical and user relevance are foundational to information science, they are insufficient for managing harmful content such as misinformation or discrimination found on the uncontrolled web. This study investigates three analytical questions: the definition of societal relevance, its practical application in search systems, and its distinction from information quality measures. By analyzing various combinations of system, user, and societal relevance, the paper explores how search outputs can be optimized for the "greater good". Although the concept remains theoretically underdeveloped, it provides a vital framework for developing value-driven search engines that prioritize ethical outcomes and societal interests over mere keyword matching.

## INTRODUCTION

Relevance is one of the fundamental concepts in information science, and there is a long history of research defining and measuring it (e.g., Buckland, 2017; Saracevic, 2016b, 2016a; White, 2010). Different types of relevance have been proposed, with the basis being topical and user relevance. However, as search systems moved from controlled collections to content from the open web, it became clear that even when a document is of topical and user relevance, it could still be harmful. Consider a document that provides misinformation about a health condition and recommends a harmful treatment. This document can be on topic and considered relevant by a user because it seems convincing. However, when we consider the greater good of society, this document is harmful and, therefore, not societally relevant in the context of this information search.

The motivation for this paper is to elaborate on the concept of societal relevance, as introduced by Jutta Haider and Olof Sundin in their 2019 book "Invisible Search and Online Search Engines" (Haider & Sundin, 2019). This concept provides a basis for a fresh discussion of what search engines and other information retrieval (IR) systems that collect documents from the open web should display. Thereby, the concept offers a way to theoretically address how to move on from showing what (often considered inappropriate) results search engines *do* display to what they *should* display.

This paper seeks to answer three analytical questions:

1. What is societal relevance, and how is it defined?
2. What would it mean if societal relevance were applied in search systems?
3. When is societal relevance really a form of relevance, and when can the problems raised in the discussion of societal relevance be solved by information quality measures?

## HAIDER AND SUNDIN'S INTRODUCTION OF SOCIETAL RELEVANCE

In short, the idea proposed by Haider and Sundin (2019) is that the classic distinction between topical relevance and user relevance, while useful in classic information retrieval settings where controlled collections are used, is not sufficient when it comes to search systems such as web search engines, which build their collections from data collected from external sources, which are not under their control. Due to the unavoidable danger of collecting documents that a society would not wish a search system to display in certain contexts, we would need a further layer of relevance, namely societal relevance, to control for cases where topical and user relevance clash with the interests of society.

It is important to note here that Haider and Sundin do not argue for excluding problematic content from search engine databases in the first place, but rather for not showing it – or showing it in lower positions – in response to search queries. Examples of content available on the web and through search engines, but often considered inappropriate, include pornography, misogyny, or results that discriminate against minorities. The most prominent collection of such cases is from Noble (2018), who shows how major web search engines return, for example, pornographic results in response to innocent queries like "black girls". Other researchers have pointed, for instance, to search results being transphobic (Graham, 2023).

According to Tefko Saracevic, topical relevance is the "relation between the subject or topic expressed in a query and topic or subject covered by information or information objects (retrieved or in the system file, or even in existence)" (Saracevic, 2016b, p. 21). So, more generally, we measure whether the topic of the result matches the topic of the request, asking whether the result is "on topic". This form of relevance is the easiest for an information retrieval system to measure, as it is a relation between the query and the document (see Mizzaro, 1997, for a hierarchy of relations that can be termed "relevance").

User relevance refers to a user's judgment of the relevance of documents returned by an information system (Saracevic, 2016b, p. 25). Even when a document is on topic, the user might not find it useful. For instance, they may already know its content, or they may not comprehend the content. User relevance is usually contrasted with system relevance, which is the system's judgment to display a result. Users, therefore, base their judgments on results the system has already deemed relevant, not on those it has deemed not relevant. The main purpose of any information retrieval system is to apply this filter: weeding out all non-relevant documents and displaying only those relevant to the user's request. Therefore, in practical information retrieval settings, we can equate topical relevance with system relevance: A user will only see documents (or surrogates of these documents) that have already qualified in terms of topical relevance, as determined by the information retrieval system.

Distinguishing between system and user relevance is sufficient as long as an information retrieval system has full control over its document collection. This applies, for instance, to a newspaper database, where articles from a particular newspaper are added daily. The newspaper has certain quality standards, and while there will be better and worse articles, the overall quality will not vary much. In contrast, a system collecting documents from a variety of external sources (particularly from the open web) will most likely face quality issues, as the quality of the documents will vary greatly, and external parties will try to influence the search system ("search engine optimization", SEO). SEO is applied not only to products and services but also to informational content, where external parties seek to influence public opinion (Lewandowski et al., 2021; Schultheiß et al., 2026), or in an academic setting, where researchers aim to increase their visibility in academic search systems and thereby increase their citation counts (Beel et al., 2010).

As soon as we have a system that collects documents from external sources outside its full control, topical and user relevance are no longer sufficient. A document may well be on topic, and the user may judge it relevant, but it might still contain disinformation or content that is considered inappropriate. This is what has been seen, for instance, in the 2016 US elections and the 2016 Brexit referendum, where large amounts of documents containing disinformation could be found in web search engines, and were selected by their users. Another simple example of topical and user relevance not being sufficient is clickbait, i.e., a document with a sensationalist headline designed to entice people to click on the link. The search system might judge the document relevant, and the user will be drawn by the headline (or snippet) and therefore select the result. However, the user might end up with misleading or even simply wrong information.

This is where Haider and Sundin's concept of societal relevance comes into play. They write,

> *while the notion of topical or subject relevance is developed and mostly given meaning in relation to books and journal articles, societal relevance also includes other aspects of the relation between information and society. When taking the question of relevance out of the traditional information science context of controlled collections to general web search engines we argue that the notion of topical or subject relevance is too limited in its scope.* (Haider & Sundin, 2019, pp. 9–10)

Thereby, the introduction of the concept of societal relevance is explicitly motivated by the need to apply relevance to search systems without full control over their collections. As relevance is always a relation (Saracevic, 2016b, p. 17), societal relevance introduces a new relation, the one between society and the document/information object. Topical relevance is the relation between a topic (as expressed in a user query) and a document, whereas user relevance is the relation between a user and a document. The problem now arising from adding the relationship between society and the document is that, while it is not easy to measure the first two relationships, measuring whether a document is relevant to society is even more challenging.

## THE SCHOLARLY RECEPTION OF SOCIETAL RELEVANCE

Since its introduction in 2019, other researchers have also used the concept of societal relevance and built on Haider and Sundin's work. Haider and Rödl (2023) point to the fact that the different forms of relevance can be contradictory to one another. Based on the term "multi-sided relevance" (Sundin et al., 2022), they argue that information ordered by Google – and we can add here: by search engines in general – inevitably occurs in differently biased ways. Interestingly, Haider and Rödl refer to relevance as determined by the system (i.e., system relevance) rather than to topical relevance, thereby motivating societal relevance as a third form of relevance.

Haider et al. (2022) introduce the term "algorithmically embodied emissions", meaning how algorithmically generated results can influence their everyday choices in the context of the climate crisis. They argue that in many searches related to that crisis, subject relevance and societal relevance are in conflict. Consequently, search engines that prioritize societal relevance could help reduce emissions by producing result sets that consider the environmental impact of choices based on those results.

Ridgway (2024) highlights a tension between societal relevance and personalization, saying that when societal relevance increases, individual relevance (here equated with personalization) decreases. Söderström and Sundin (2024) point out that societal relevance can conflict not only with other types of relevance but also, by aligning results with legislation and established science, run counter to users' wishes. Also, they discuss how actors, in their examples, grey-zone online pharmacies, circumvent societal relevance by using search engine optimization (SEO) techniques, trying to make search engines believe their websites are trustworthy and meet the criteria for societal relevance. Gorichanaz (2023) argues that search engines like Google should promote "good thinking" in their results, so that they do not force a shared understanding between users but valid facts to allow for informed discussions. He also puts societal relevance in the context of legislation, namely the European General Data Protection Regulation (GDPR), with the "Right to be forgotten" (RTBF). With a successful RTBF request, a person can force a search engine not to display particular results in response to a search query for that person's name. For instance, for someone who committed a crime and has served their prison sentence, search results for that crime will be demoted. Thereby, a search engine will demote results that are on topic and could be relevant to a user, but do not have societal relevance (anymore). Similarly, Munton (2022) argues that "On epistemic grounds, some information should be returned in response to a given inquiry, even if it does not serve the goals of the user of the search engine at that time." (p. 2)

Societal relevance has also been discussed in the context of data voids (Golebiewski & Boyd, 2018), i.e., cases where there are no documents of sufficient quality available on the web and, therefore, a search engine outputs the documents that are most relevant in the sense of topical and user relevance, but not relevant in the sense of societal relevance (Norocel & Lewandowski, 2023). Other researchers (P et al. ,2024) argue that societal relevance can be used to exclude potentially pernicious web pages, thereby shifting the concept from a relevance (a relation) to an information-quality measure (see section "Is it really a form of relevance?").

## DEFINING SOCIETAL RELEVANCE IN THE CONTEXT OF SEARCH OUTPUTS

Apart from a general discussion of societal relevance, it is important to consider what it would mean to apply societal relevance to search outputs, e.g., from a search engine or an AI chatbot. When we consider user relevance and system relevance only, the goal of an information retrieval system (and what is measured in IR testing; see Saracevic, 2016b, p. 20) is to match the two types of relevance. This would mean, in an ideal case, the system outputs only results the user finds relevant, or at least outputs all relevant items before any not relevant ones (Buckland, 2016).

Now, when the interests of society come into play as a third factor, we first need to ask: which society? It surely makes a difference whether we talk about a democratic or an authoritarian society, and we cannot assume there is a single measure of societal relevance that applies to all. However, putting this problem in the spotlight can help us promote the idea that search engine companies should take responsibility for their results and that those results are not necessarily a mirror of users' beliefs.

Noble (2018) argues that groups should be in control of what the results about them look like. This is also highly problematic, as any group could promote results in their favor and demote results critical of them. Tripodi (2022) warns of the dangers of "equal coverage", or in information retrieval terms, strict result diversity. In terms of societal relevance, simply providing all available opinions on a topic will not lead anywhere, as it would only promote obscure, fact-free opinions.

We can distinguish between two types of searches:

(1) Someone is searching for information on a topic, and there are documents deemed societally relevant and others deemed not. In this case, and if there is no data void (see above), the search system just needs to rank the societally relevant documents first, as described in the last section.
(2) Someone is searching for a topic whose query itself indicates they are seeking information that is neither societally relevant nor appropriate. Gorichanaz (2023) provides the example of someone searching for how to commit suicide. The question here is whether a search engine should display no matching results at all and/or provide information boxes with specific suicide-prevention information, as described, e.g., by Haim et al., (2017).

## WHAT WOULD IT MEAN TO APPLY SOCIETAL RELEVANCE IN SEARCH SYSTEMS?

On a more practical level, introducing a societal relevance metric to measure a search system's output further complicates the problem. If we just use binary judgments (relevant, not relevant), introducing a third type of relevance leads to eight (2 x 2 x 2) conditions (see Table 1), assuming only responses that are displayed by the search system (i.e., leaving all other documents/responses). The following discussion applies to any type of search output, for instance, ranked lists of documents or AI answers. I will, therefore, use the term "response" instead of "document", "information object", or "answer".

In the following discussion, we assume an information retrieval system that can output topically relevant results. Therefore, we can equate topical relevance with system relevance. Only this distinction will allow us to predict which result a user will select/consider. (Even in the case of the AI answer – which is obviously not clickable – a user will make the decision of whether to consider it or not.) Another limitation of this discussion is that it focuses only on binary relevance for clarity, although in productive settings, graded relevance is the standard.

Here, we are interested in the effect of the different combinations of relevance judgments. We can easily see that the overall effect is positive only in three cases: (1) when the topical, user, and societal relevance are judged as positive, (2) when the topical, user, and societal relevance are judged as negative, (3) when the output is of neither topical nor societal relevance but of user relevance.

In the case of positive judgments only, the result will be on-topic, the user will appreciate it, and it will benefit society. In the negative case, neither is the case, and it is, therefore, obvious that the response will not be considered further.

More interesting are the cases in between, i.e., where at least one type of relevance is considered fulfilled. First, let's consider topically relevant results:

When a response fulfills the criterion of topical relevance but not the other two forms of relevance, this response creates ballast in the result presentation: it is obvious from its not being on topic, and users judged it as not relevant, yet it is still presented by the system. Of course, such a response poses no danger, as it is not considered anyway. It may only lead to a user having to wade through a longer result list or, in the case of an AI answer, not considering the answer at all, as not more than one such answer is provided.

When a result fulfills the criteria of topical relevance and user relevance, this also leads to a negative effect, although a different one: Here, the user would consider that response, even though it does not provide societal relevance, i.e., it could provide misleading, outright wrong, or otherwise unwanted information.

When a result is topically relevant and societally relevant, but the user sees it as not relevant, the negative outcome is that the user might not consider it relevant. However, this only holds under the assumption that outside parties (in this case, the system and whoever determined what is societally relevant) know better than the user what would be relevant to them. While we can imagine many scenarios where this is the case, it could also be the case that the user rightfully considers an otherwise relevant item not relevant.

Next, consider the topically not relevant results.

When a result is topically not relevant, and the user also considers it not relevant, but it is of societal relevance, the user will miss an otherwise relevant result because they will not see it, as it is not displayed by the system. The same goes for the combination topically not relevant, fulfils user relevance, fulfils societal relevance. Note that it does not matter whether the user considers the document relevant, as it will not be displayed by the system.

The last case (topical relevance: no, user relevance: yes, societal relevance: no) is noteworthy because the positive effect of this constellation might be counterintuitive: why should it be possible when a response that a user will judge as not relevant is not displayed? As the document is not societally relevant, it is positive that the system will not output it (because of its lack of topical relevance). The user will not see the response they would consider relevant, but that could lead them to misinformation, believing it is relevant to them.

We can summarize that a search system should always at least prioritize responses that fulfill all three forms of relevance. However, the question remains: how should a system handle queries for which no such documents can be identified?

| Topical/system | User relevance | Societal relevance | Effect |
| --- | --- | --- | --- |
| relevant | relevant | relevant | Positive |
| relevant | not relevant | not relevant | Negative: fallout in the sense of precision/recall |
| relevant | relevant | not relevant | Negative: user selects a result that may be false or misleading |
| relevant | not relevant | relevant | Can be negative: user might miss a relevant result |
| not relevant | relevant | relevant | Negative: User will miss the result because the system will not show it |
| not relevant | not relevant | relevant | Negative: User will miss the result because the system will not show it |
| not relevant | relevant | not relevant | Positive: User will not see the result they would erroneously regard as relevant |
| not relevant | not relevant | not relevant | Positive: result will not be shown |

**Table 1. Effects of combinations of topical/system relevance, user relevance, and societal relevance**

## IS IT REALLY A FORM OF RELEVANCE?

It might seem confusing to discuss whether societal relevance is, in fact, a form of relevance *after* presenting it as a third form of relevance. However, the discussion so far has focused on the consequences of introducing societal relevance, whereas this section asks how to address the problem posed by non-societally relevant documents in a search system's data store.

The easiest solution would be not to index such documents in the first place, thereby avoiding the negative consequences discussed in the last section. This would mean that search engines would be much more selective and move from an approach of more or less "indexing everything" (minus spam; see Lewandowski, 2023, pp. 39–40) to curating document collections. It is obvious that this would be at the expense of comprehensiveness. Furthermore, a search engine company following this approach would have to define clear rules on what to include in its collection and what to exclude. One can imagine that this would lead to extensive discussions of censorship, especially when a search engine with a considerable market share, especially Google with its dominant position in the search engine market (Statcounter, 2025), would do so. However, we should keep in mind that while search engines usually do not have explicit editorial criteria for including documents in their data stores, they do have implicit editorial criteria in their ranking functions. In practice, this leads to very different results depending on which search engine a user prefers. Overlap between leading search engines is lower than often expected (Yagci et al., 2022).

This outlined solution would not require societal relevance, but societal appropriateness, moving the concept into a different category: from relevance (a relation to something) to information quality (a property of an information object independent of a searcher's context). The problem with this solution is that a document either meets the minimum criteria for inclusion (a threshold for information quality) or does not.

However, it could well be the case that a document is appropriate for some queries but not for others. Consider a document that presents quotes from Donald Trump on how to treat women. It is a difference whether it is shown in response to the query "How should you treat women?" or "What did Donald Trump say about how you can treat women when you're rich and famous?" Considering this distinction, it makes sense to introduce societal relevance as a third form of relevance, as it is about a relation (between the document and its societal appropriateness *in the context of the query*). A solution, then, is for a search engine to consider societal relevance and downrank documents deemed inappropriate in the context of the query. Although unlikely, a user could then still move from result page to result page to find these documents in their low positions. However, a search engine could also not rank these documents at all in response to particular queries, as in the example of the "Right to be forgotten" discussed above. This would make it impossible for a user to find them by clicking through to lower-ranked results. However, a user might circumvent the search engine's decision not to show these results by simply changing the query to something that appears to be an appropriate request for that document (if they know about it; see the example above).

While there is no solution that solves the problems outlined, the discussion also helps us classify research that audits search engine outputs and, from these outputs, derives what should (or should not) be shown in (top) search results relative to what is actually shown. Unfortunately, much of that research simply states that some results should simply not be shown (e.g., Noble, 2018), and does not distinguish whether the results are societally relevant in the context of the query. Further discussion of societal relevance could help develop value-driven search engines.

## CONCLUSION

In conclusion, while Haider and Sundin introduced the concept of societal relevance in 2019, it still lacks a formal definition and remains largely underdeveloped in its theoretical framework. Despite this, it is an undeniably relevant (!) concept for information science, offering a necessary lens for search systems that no longer operate with fully controlled collections. By exploring the practical consequences of its application, this paper contributes a step toward establishing societal relevance in both theory and practice. The discussion remains caught in a tension between using ranking functions to demote inappropriate results and the more extreme measure of excluding them from collections entirely. This choice forces a shift from simple relevance to "societal appropriateness," raising concerns about curation and censorship. Ultimately, this analysis raises more questions than it provides definitive answers, particularly regarding whose societal values should be prioritized. Future information science research must continue to grapple with these complexities to determine if the concept has a fruitful and sustainable future in the development of value-driven search engines and other search systems.

## AUTHOR ATTRIBUTION

Author: Conceptualization, investigation, methodology, project administration, resources, software, validation, writing – original draft, and writing – review and editing.

## GENERATIVE AI USE

We employed Grammarly and Google Gemini for the following purpose: language editing. The author assumes all responsibility for the content of this submission.